\documentclass[conference]{IEEEtran}
\IEEEoverridecommandlockouts

\usepackage{amsthm}

 \newtheorem{theorem}{Theorem}
 
 \newtheorem{lemma}{Lemma}

 \newtheorem{definition}[theorem]{Definition}

\usepackage{amssymb}
\usepackage{cite}

\ifCLASSINFOpdf
\usepackage[pdftex]{graphicx}
\graphicspath{{./Figures/}}
\else
\fi

\usepackage[cmex10]{amsmath}
\usepackage{algorithmic}
\usepackage{array}
\usepackage{url}
\usepackage{mathrsfs}

\usepackage[dvipsnames,svgnames]{xcolor}
\usepackage{bbm}

\begin{document}

\title{Spatio-Temporal Visualization of Interdependent Battery Bus Transit and Power Distribution Systems
}

\author{\IEEEauthorblockN{Avishan Bagherinezhad\IEEEauthorrefmark{1}, Michael Young\IEEEauthorrefmark{2}, Bei Wang\IEEEauthorrefmark{2}, Masood Parvania\IEEEauthorrefmark{1}}
\IEEEauthorblockA{\IEEEauthorrefmark{1}Department of Electrical and Computer Engineering, University of Utah, Salt Lake City, UT 84112\\
\IEEEauthorrefmark{2}School of Computing, University of Utah, Salt Lake City, UT 84112 \\
Emails: \{avi.bagherinezhadsowmesaraee, masood.parvania\}@utah.edu, myoung@cs.utah.edu, beiwang@sci.utah.edu}
}

\newcommand{\force}{\mbox{$\Vdash$}}

\newcommand{\ghat}{\mbox{$\bm \hat g$}}

\newcommand{\bzero}{\mbox{${\bm 0}$}}

\newcommand{\va}{\mbox{${\mathbf a}$}}
\newcommand{\vah}{\mbox{${\mathbf \hat a}$}}
\newcommand{\vat}{\mbox{${\mathbf \tilde a}$}}
\newcommand{\vb}{\mbox{${\mathbf b}$}}
\newcommand{\vd}{\mbox{${\mathbf d}$}}
\newcommand{\rh}{\mbox{${\hat r}$}}
\newcommand{\Itl}{\mbox{${\tilde I}$}}

\newcommand{\vxt}{\mbox{${\mathbf \tilde x}$}}
\newcommand{\vh}{\mbox{${\mathbf h}$}}
\newcommand{\vhh}{\mbox{${\mathbf \hat h}$}}
\newcommand{\ve}{\mbox{${\mathbf e}$}}
\newcommand{\vg}{\mbox{${\mathbf g}$}}
\newcommand{\vgh}{\mbox{${\mathbf \hat g}$}}
\newcommand{\vp}{\mbox{${\mathbf p}$}}
\newcommand{\vph}{\mbox{${\mathbf \hat p}$}}
\newcommand{\vq}{\mbox{${\mathbf q}$}}
\newcommand{\vt}{\mbox{${\mathbf t}$}}
\newcommand{\vw}{\mbox{${\mathbf w}$}}
\newcommand{\vwh}{\mbox{${\mathbf \hat w}$}}
\newcommand{\wh}{\mbox{${\hat w}$}}
\newcommand{\vwt}{\mbox{${\mathbf \tilde w}$}}
\newcommand{\wt}{\mbox{${\tilde w}$}}
\newcommand{\vs}{\mbox{${\mathbf s}$}}
\newcommand{\vsh}{\mbox{${\mathbf \hat s}$}}
\newcommand{\vst}{\mbox{${\mathbf \tilde s}$}}
\newcommand{\vr}{\mbox{${\mathbf r}$}}
\newcommand{\vx}{\mbox{${\mathbf x}$}}
\newcommand{\vv}{\mbox{${\mathbf v}$}}
\newcommand{\vu}{\mbox{${\mathbf u}$}}
\newcommand{\vy}{\mbox{${\mathbf y}$}}
\newcommand{\vz}{\mbox{${\mathbf z}$}}
\newcommand{\vn}{\mbox{${\mathbf n}$}}
\newcommand{\vnt}{\mbox{${\mathbf \tilde n}$}}
\newcommand{\vzero}{\mbox{${\mathbf 0}$}}
\newcommand{\vone}{\mbox{${\mathbf 1}$}}

\newcommand{\mA}{\mbox{{$\mathbf A$}}}
\newcommand{\mAh}{\mbox{${\mathbf \hat A}$}}
\newcommand{\mAt}{\mbox{$\mathbf \tilde A$}}
\newcommand{\mB}{\mbox{${\mathbf B}$}}
\newcommand{\mBh}{\mbox{${\mathbf \hat B}$}}
\newcommand{\mC}{\mbox{{$\mathbf C$}}}
\newcommand{\mCh}{\mbox{${\mathbf \hat C}$}}
\newcommand{\mD}{\mbox{{$\mathbf D$}}}
\newcommand{\mDt}{\mbox{$\mathbf \tilde D$}}
\newcommand{\mE}{\mbox{{$\mathbf E$}}}
\newcommand{\mG}{\mbox{{$\mathbf G$}}}
\newcommand{\mF}{\mbox{{$\mathbf F$}}}
\newcommand{\mH}{\mbox{{$\mathbf H$}}}
\newcommand{\mHb}{\mbox{${\mathbf \bar H}$}}
\newcommand{\mI}{\mbox{{$\mathbf I$}}}
\newcommand{\mIh}{\mbox{${\mathbf \hat I}$}}
\newcommand{\mN}{\mbox{{$\mathbf N$}}}
\newcommand{\mM}{\mbox{{$\mathbf M$}}}
\newcommand{\mMh}{\mbox{{$\mathbf \hat M$}}}
\newcommand{\mP}{\mbox{${\mathbf P}$}}
\newcommand{\mQ}{\mbox{${\mathbf Q}$}}
\newcommand{\mR}{\mbox{${\mathbf R}$}}
\newcommand{\mRh}{\mbox{${\mathbf {\hat {R}}}$}}
\newcommand{\mRt}{\mbox{${\mathbf \tilde R}$}}
\newcommand{\mS}{\mbox{${\mathbf S}$}}
\newcommand{\mSb}{\mbox{${\mathbf \bar S}$}}
\newcommand{\mSh}{\mbox{${\mathbf \hat S}$}}
\newcommand{\mSt}{\mbox{${\mathbf \tilde S}$}}
\newcommand{\mT}{\mbox{${\mathbf T}$}}
\newcommand{\mU}{\mbox{${\mathbf U}$}}
\newcommand{\mUh}{\mbox{${\mathbf \hat U}$}}
\newcommand{\mV}{\mbox{${\mathbf V}$}}
\newcommand{\mVh}{\mbox{${\mathbf \hat V}$}}
\newcommand{\mW}{\mbox{${\mathbf W}$}}
\newcommand{\mWh}{\mbox{${\mathbf \hat W}$}}
\newcommand{\mWt}{\mbox{${\mathbf \tilde W}$}}
\newcommand{\mX}{\mbox{${\mathbf X}$}}
\newcommand{\mY}{\mbox{${\mathbf Y}$}}
\newcommand{\mZ}{\mbox{${\mathbf Z}$}}


\newcommand{\ga}{\alpha}
\newcommand{\gb}{\beta}
\newcommand{\grg}{\gamma}
\newcommand{\gd}{\delta}
\newcommand{\gre}{\varepsilon}
\newcommand{\gep}{\epsilon}
\newcommand{\gz}{\zeta}
\newcommand{\gzh}{\mbox{$ \hat \zeta$}}
\newcommand{\gh}{\eta}
\newcommand{\gth}{\theta}
\newcommand{\gi}{iota}
\newcommand{\gk}{\kappa}
\newcommand{\gl}{\lambda}
\newcommand{\gm}{\mu}
\newcommand{\gn}{\nu}
\newcommand{\gx}{\xi}
\newcommand{\gp}{\pi}
\newcommand{\gph}{\phi}
\newcommand{\gr}{\rho}
\newcommand{\gs}{\sigma}
\newcommand{\gsh}{\hat \sigma}
\newcommand{\gt}{\tau}
\newcommand{\gu}{\upsilon}
\newcommand{\gf}{\varphi}
\newcommand{\gc}{\chi}
\newcommand{\go}{\omega}


\newcommand{\gG}{\Gamma}
\newcommand{\gD}{\Delta}
\newcommand{\gTh}{\Theta}
\newcommand{\gL}{\Lambda}
\newcommand{\gX}{\Xi}
\newcommand{\gP}{\Pi}
\newcommand{\gS}{\Sigma}
\newcommand{\gU}{\Upsilon}
\newcommand{\gF}{\Phi}
\newcommand{\gO}{\Omega}


\def\bm#1{\mbox{\boldmath $#1$}}
\newcommand{\vga}{\mbox{$\bm \alpha$}}
\newcommand{\vgb}{\mbox{$\bm \beta$}}
\newcommand{\vgd}{\mbox{$\bm \delta$}}
\newcommand{\vge}{\mbox{$\bm \epsilon$}}
\newcommand{\vgl}{\mbox{$\bm \lambda$}}
\newcommand{\vgm}{\mbox{$\bm \mu$}}
\newcommand{\vgr}{\mbox{$\bm \rho$}}
\newcommand{\vgn}{\mbox{$\bm \nu$}}
\newcommand{\vgp}{\mbox{$\bm \pi$}}
\newcommand{\vgrh}{\mbox{$\bm \hat \rho$}}
\newcommand{\vgrt}{\mbox{$\bm {\tilde \rho}$}}

\newcommand{\vgt}{\mbox{$\bm \gt$}}
\newcommand{\vgth}{\mbox{$\bm {\hat \tau}$}}
\newcommand{\vgtt}{\mbox{$\bm {\tilde \tau}$}}
\newcommand{\vpsi}{\mbox{$\bm \psi$}}
\newcommand{\vphi}{\mbox{$\bm \phi$}}
\newcommand{\vxi}{\mbox{$\bm \xi$}}
\newcommand{\vth}{\mbox{$\bm \theta$}}
\newcommand{\vthh}{\mbox{$\bm {\hat \theta}$}}

\newcommand{\mgG}{\mbox{$\bm \Gamma$}}
\newcommand{\mgGh}{\mbox{$\hat {\bm \Gamma}$}}
\newcommand{\mgD}{\mbox{$\bm \Delta$}}
\newcommand{\mgU}{\mbox{$\bm \Upsilon$}}
\newcommand{\mgL}{\mbox{$\bm \Lambda$}}
\newcommand{\mPsi}{\mbox{$\bm \Psi$}}
\newcommand{\mgX}{\mbox{$\bm \Xi$}}
\newcommand{\mgS}{\mbox{$\bm \Sigma$}}

\newcommand{\oA}{{\open A}}
\newcommand{\oC}{{\open C}}
\newcommand{\oF}{{\open F}}
\newcommand{\oN}{{\open N}}
\newcommand{\oP}{{\open P}}
\newcommand{\oQ}{{\open Q}}
\newcommand{\oR}{{\open R}}
\newcommand{\oZ}{{\open Z}}


\newcommand{\Nu}{{\cal V}}
\newcommand{\cA}{{\cal A}}
\newcommand{\cB}{{\cal B}}
\newcommand{\cC}{{\cal C}}
\newcommand{\cD}{{\cal D}}
\newcommand{\cF}{{\cal F}}
\newcommand{\cH}{{\cal H}}
\newcommand{\cK}{{\cal K}}
\newcommand{\cI}{{\cal I}}
\newcommand{\cL}{{\cal L}}
\newcommand{\cM}{{\cal M}}
\newcommand{\cN}{{\cal N}}
\newcommand{\cO}{{\cal O}}
\newcommand{\cP}{{\cal P}}
\newcommand{\cR}{{\cal R}}
\newcommand{\cS}{{\cal S}}
\newcommand{\cU}{{\cal U}}
\newcommand{\cV}{{\cal V}}
\newcommand{\cT}{{\cal T}}
\newcommand{\cX}{{\cal X}}

\newcommand{\rH}{^{*}}
\newcommand{\rT}{^{ \raisebox{1.2pt}{$\rm \scriptstyle T$}}}
\newcommand{\rF}{_{ \raisebox{-1pt}{$\rm \scriptstyle F$}}}
\newcommand{\rE}{{\rm E}}

\newcommand{\dom}{\hbox{dom}}
\newcommand{\rng}{\hbox{rng}}
\newcommand{\Span}{\hbox{span}}
\newcommand{\Ker}{\hbox{Ker}}
\newcommand{\On}{\hbox{On}}
\newcommand{\otp}{\hbox{otp}}
\newcommand{\ZFC}{\hbox{ZFC}}
\def\Re{\ensuremath{\hbox{Re}}}
\def\Im{\ensuremath{\hbox{Im}}}
\newcommand{\SNR}{\ensuremath{\hbox{SNR}}}
\newcommand{\CRB}{\ensuremath{\hbox{CRB}}}
\newcommand{\diag}{\ensuremath{\hbox{diag}}}
\newcommand{\trace}{\ensuremath{\hbox{tr}}}

\newcommand{\dlot}{\mbox{$\delta^1_3$}}
\newcommand{\Dlot}{\mbox{$\Delta^1_3$}}
\newcommand{\Dlof}{\mbox{$\Delta^1_4$}}
\newcommand{\dlof}{\mbox{$\delta^1_4$}}
\newcommand{\bP}{\mbox{$\mathbf{P}$}}
\newcommand{\Pot}{\mbox{$\Pi^1_2$}}
\newcommand{\Sot}{\mbox{$\Sigma^1_2$}}
\newcommand{\gDot}{\mbox{$\gD^1_2$}}

\newcommand{\Potr}{\mbox{$\Pi^1_3$}}
\newcommand{\Sotr}{\mbox{$\Sigma^1_3$}}
\newcommand{\gDotr}{\mbox{$\gD^1_3$}}

\newcommand{\Pofr}{\mbox{$\Pi^1_4$}}
\newcommand{\Sofr}{\mbox{$\Sigma^1_4$}}
\newcommand{\Dofr}{\mbox{$\gD^1_4$}}

\newcommand{\Sa}{\mbox{$S_{\ga}$}}
\newcommand{\Qk}{\mbox{$Q_{\gk}$}}
\newcommand{\Ca}{\mbox{$C_{\ga}$}}

\newcommand{\gkp}{\mbox{$\gk^+$}}
\newcommand{\aron}{ Aronszajn }

\newcommand{\sqkp}{\mbox{$\Box_{\gk}$}}
\newcommand{\dkp}{\mbox{$\Diamond_{\gk^{+}}$}}
\newcommand{\sqsqnce}
{\mbox{\\ $\ < \Ca \mid \ga < \gkp \ \ \wedge \ \ \lim \ga >$ \ \ }}
\newcommand{\dsqnce}{\mbox{$<S_{\ga} \mid \ga < \gkp >$}}



\newcommand{\beq}{\begin{equation}}
\newcommand{\eeq}{\end{equation}}
\newcommand{\bea}{\begin{array}}
\newcommand{\ena}{\end{array}}
\newcommand{\bds}{\begin {itemize}}
\newcommand{\eds}{\end {itemize}}
\newcommand{\bdf}{\begin{definition}}
\newcommand{\edf}{\end{definition}}
\newcommand{\blm}{\begin{lemma}}
\newcommand{\elm}{\end{lemma}}
\newcommand{\bthm}{\begin{theorem}}
\newcommand{\ethm}{\end{theorem}}
\newcommand{\bprp}{\begin{prop}}
\newcommand{\eprp}{\end{prop}}
\newcommand{\bcl}{\begin{claim}}
\newcommand{\ecl}{\end{claim}}
\newcommand{\bcr}{\begin{coro}}
\newcommand{\ecr}{\end{coro}}
\newcommand{\bquest}{\begin{question}}
\newcommand{\equest}{\end{question}}

\newcommand{\rarrow}{{\rightarrow}}
\newcommand{\Rarrow}{{\Rightarrow}}
\newcommand{\larrow}{{\larrow}}
\newcommand{\Larrow}{{\Leftarrow}}
\newcommand{\restrict}{{\upharpoonright}}
\newcommand{\nin}{{\not \in}}



\newcommand{\ie}{\hbox{i.e.}}
\newcommand{\eg}{\hbox{e.g.}}

\maketitle

\begin{abstract}
The high penetration of transportation electrification and its associated charging requirements magnify the interdependency of the transportation and power distribution systems. 
The emergent interdependency requires that system operators fully understand the status of both systems.
To this end, a visualization tool is presented to illustrate the interdependency of battery bus transit and power distribution systems and the associated components.
The tool aims at monitoring components from both systems, such as the locations of electric buses, the state of charge of batteries, the price of electricity, voltage, current, and active/reactive power flow. 
The results showcase the success of the visualization tool in monitoring the bus transit and power distribution components to determine a reliable cost-effective scheme for spatio-temporal charging of electric buses.
\end{abstract}

\IEEEpeerreviewmaketitle

\section{Introduction}
\IEEEPARstart{T}{he} electrification of the transportation systems offers compelling potential to reduce greenhouse gas emissions and to combat climate change.
In the United States, the transportation sector accounted for 28\% of total greenhouse gas emissions in 2018 \cite{epa}.
For this reason, transportation electrification is recognized as a proactive solution to emergent environmental challenges.
The paradigm shift toward adopting zero-emission battery electric buses (BEBs) plays a crucial role in eliminating tailpipe emissions and reducing the negative impacts of fossil fuel-based public transportation. 
According to the International Energy Agency, the number of on-road electric buses will reach 1.5 million by 2030 \cite{bunsen2018global}. 

An adequate charging infrastructure is crucial in maintaining the reliability of the battery bus transit (or transit for short) system operation.
The BEBs have the opportunity to charge their batteries at either a charging depot or an on-route charging station.
For depot charging, BEBs charge their batteries over night for 3 to 7 hours, with 50-120 kW chargers. 
Such BEBs require large battery packs to maintain the daily transit schedule without charging. 
In comparison, for on-route charging, BEBs require smaller battery packs, which are regularly charged at on-route charging stations to maintain the transit schedule~\cite{prohaska2016fast}.  
On-route charging stations require a fast charging infrastructure (i.e.,~500 kW chargers) to provide charging opportunities during the dwell time of BEBs (5 to 10 minutes) at the bus stops.
The depot charging takes advantage of the nighttime off-peak electricity tariff, and on-route charging benefits transit and power distribution systems by enabling both spatial and temporal flexibility in BEBs' charging demand. 

In the traditional urban structure, bus transit system offers shared passenger transport services across the city, whereas the power distribution system supplies individual customer electricity demands. 
However, current trends in adopting battery bus transit system and the supporting policies cast light on the synergy of both systems. 
Although the high penetration of battery bus transit system has environmental merits, minimal coordination and planning pose new challenges~\cite{Energy}. 
The uncoordinated operation of both systems may violate the time-tables of the BEBs, and jeopardize the reliability of the power distribution system in terms of frequency, voltage, and line congestion~\cite{palomino2019advanced}.

The decision-making process based on the heterogeneous data that arise from battery bus transit system and power distribution system can be complex and intricate. 
Operators from both systems can benefit from data visualization, which conveys information in an interactive, intuitive, and informative way. 
In this paper, we present a visualization tool that captures the interdependency of the battery bus transit and power distribution systems, referred to as the \emph{interdependent systems} for short. 
In Section \ref{sec.interdependent}, we discuss the interdependent systems,  their network models, and opportunities and challenges. In Section \ref{sec.literature}, we present related works.
In Section \ref{sec.vis}, we describe the visualization techniques for modeling such an interdependent system. 
We discuss applications of the visualization tool and conclude the paper in Section \ref{sec.conclusion}.

\section{Interdependent Battery Bus Transit and Power Distribution Systems}
\label{sec.interdependent}
Bus transit is a critical infrastructure system for serving passengers in urban environments.
The ultimate goal of a transit system operator is to improve the quality of service while minimizing the total operation cost.
Electrification of the bus transit system imposes two additional costs: energy and demand charge.
The energy cost is associated with the consumed energy (kWh) based on different electricity tariff structures, whereas the demand charge cost is calculated based on the maximum required power (kW) in 15 minutes or hourly intervals during a month \cite{fitzgerald2017evgo}.

Each BEB's schedule is fixed in terms of its traveling time, distance, and the roads traveled during each trip: a BEB stops at a set of bus stops at specific times; and some of these stops are equipped with on-route chargers.
The transit system operator monitors the location and the state of charge (SOC) of BEB batteries to schedule the charging events so that the system constraints are not violated. 
The bus stops with charging infrastructure draw power from the power distribution system to cater to the spatio-temporal charging demand.
The sustainability of BEBs' transit schedule relies on the adequacy of the charging infrastructure, which highlights the interdependency of the bus transit and power distribution systems. 
Therefore, transit operators need to monitor the power distribution system constraints to ensure reliable and cost-effective operation of transit system. 
The same principle applies to power system operators.

\subsection{Opportunities}
Adopting independently-developed visualization tools for battery bus transit and power distribution systems is inadequate due to their inevitable interdependency. 

The complex data in these coupled systems call for integrated and multi-layer visualization, which will offer unique insights into the underlying data and help operators of both systems comprehend, communicate, and make decisions.  

The bus transit system is subject to both temporal and spatial changes.
A multi-layer visualization will provide information regarding the number of BEBs on the road and at stations, locations of BEBs, SOC of BEB batteries, bus stops within BEBs' trips, and available power capacity at bus stops with charging infrastructure over time. 
Moreover, this visualization will allow the power distribution system operators to monitor the active and reactive power flow, and to observe the impact of spatio-temporal charging demand of BEBs on the system.

Although the decision variables of the power distribution and bus transit system appear to be disjoint, the decisions  on one system have a substantial impact on the other.
Independent scheduling of the bus transit system may result in a power outage by violating power distribution system constraints, which impacts the reliability and operation costs of the transit system.
For this reason, coupling a co-optimization model (e.g., \cite{bagherinezhad2020spatio}) with visualization is beneficial for  quickly assessing changes in the system and applying corrective measures.
The co-optimization ensures minimum operation costs of both systems and the visualization reveals their interdependency.
Therefore, the list of data to be visualized should include:  
\begin{itemize}
    \item Spatio-temporal locations of BEBs in the transit system;
    \item SOC of BEB batteries and minimum and maximum energy thresholds in the transit system;
    \item Location of bus stops with on-route chargers; 
    \item Spatio-temporal charging demand of BEBs; 
    \item Active and reactive load at power distribution system;
    \item Power distribution's active and reactive power flow; 
    \item Current flow and maximum capacity of the power lines; 
    \item Voltage level of the power distribution system nodes, and maximum and minimum voltage thresholds;
    \item Electricity price paid by interdependent systems.
\end{itemize}

The successful operation of the interdependent systems is dependent on assessing the data in this list meticulously and making decisions with corrective measures accordingly.
The intuitive and simple representation of the power distribution data gives insights to transit system operators to adapt the charging scheme of BEBs spatially and temporally to maintain transit system time-table and power distribution system constraints and to reduce energy and demand charge costs.

\subsection{Challenges}
The efficacy of independent scheduling and the operation of the interdependent systems is questionable due to their intertwined infrastructures.
For instance, scheduling BEB charging events during off-peak hours minimizes the energy cost; however, the emergent peak in charging demand may result in congestion in power lines and violation of voltage limits.
Therefore, a coordinated co-optimization strategy is crucial to secure the reliable operation of both systems. 
Further, the efforts and resources to visualize the complex data associated with the operation of the interdependent systems would be squandered if the data were obsolete by the time they are presented to the operators. 
For this reason, robust visualization tools should be able to ingest and process live stream data to help decision makers trace trends and patterns to make timely decisions.

\subsection{Bus Transit and Power Distribution Model}
The bus transit and power distribution systems are modeled by a graph $G=(\mathcal{N}, \mathcal{L})$, where $\mathcal{N}$ and $\mathcal{L}$ represent the sets of nodes and links (edges) that connect the nodes, respectively. 
The bus transit system comprises bus stops, roads, and BEBs. 
The battery bus transit system is defined as a subset of the transit system, in which a subset of existing bus stops equipped with charging infrastructure defines the set of nodes; and roads that connect such stops define the set of links. 
The power distribution system attributes include nodes, lines, inflexible and flexible electric loads, substation, and complex impedance and admittance. 
The current, active power flow, and reactive power flow are attributes of the lines, and the charging profile of charging stations, voltage magnitude, and angle are attributes of nodes in the power distribution system.

\begin{figure}[b] 
	\centering
	\includegraphics[width=\linewidth]{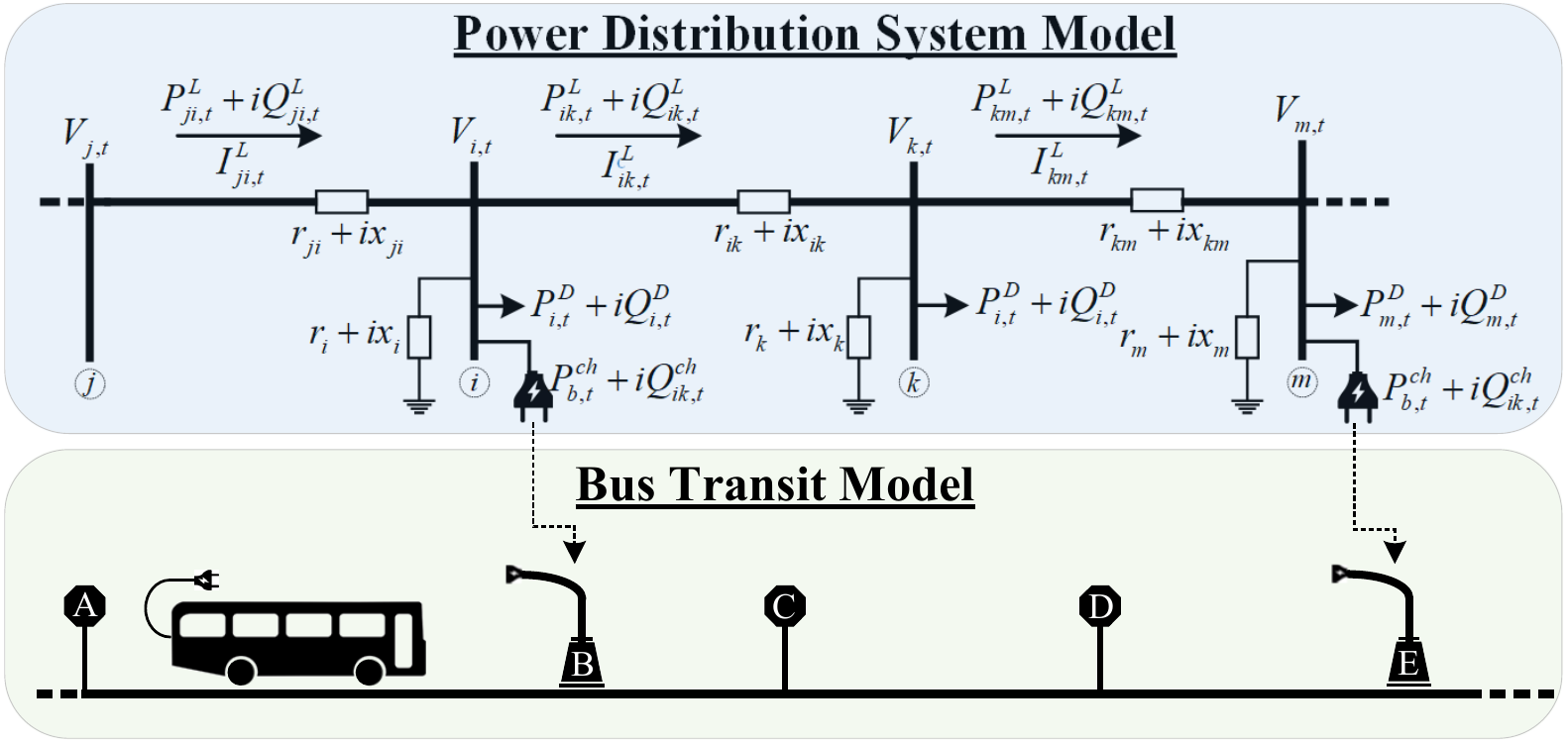}
  \vspace{-15pt}
	\caption{Interdependent bus transit and power distribution systems.}
	\label{fig.PowerTrans} 
\end{figure}
In Fig.~\ref{fig.PowerTrans}, we depict the interdependent bus transit and power distribution systems.
In Fig.~\ref{fig.PowerTrans}, BEB passes by bus stops $A$ to $D$, in which only bus stops $B$ and $E$ are equipped with charging infrastructure. Given the availability of the on-route charging station, BEB has the opportunity to choose between bus stops $B$ and $E$ for charging the batteries.
The BEB charging requirements at charging stations $B$ and $E$ in the transit system are considered as flexible load at nodes $i$ and $m$ in the power distribution system. 
The spatio-temporal charging requirement of BEBs in different bus stops with charging infrastructure makes the two systems interdependent.

\section{Related Work}
\label{sec.literature}
A number of works have focused on the visualization of transportation systems, power systems, and multi-layer networks~\cite{ZengFuArisona2014, ChenGuoWang2015,AndrienkoAndrienkoChen2017,AhmedMiller2007,KurkcuMirandaOzbay2017,OverbyeWeber2000,HockMcGuiness2018,McGeeGhoniemMelancon2019,InterdonatoMagnaniPerna2020}.
In addition, commercial software tools such as \textit{MercuryGate} and \textit{Oracle} that handle operation planning and fleet management for transportation systems~\cite{mercury, oracle},  \textit{PowerFactory}, \textit{PowerWorld}, and \textit{Cyme} for power systems provide visualization components as well~\cite{digsilent,powerworld,cyme}. 
Zeng et al.~\cite{ZengFuArisona2014} captured and visualized the mobility of the passengers in the public transportation system by utilizing passengers' radio-frequency identification card data.
Their model visualized mobility, geographical and temporal information by  isotime flow map and origin destination (OD) pair journey models~\cite{ZengFuArisona2014}.

Chen el al.~\cite{ChenGuoWang2015} and Andrienko et al.~\cite{AndrienkoAndrienkoChen2017} provided, respectively, surveys on the data processing and visualization techniques that illustrate spatio-temporal and multi-variable characteristics of traffic data and discussed the impact of transportation system on other infrastructures.
Ahmed et al.~\cite{AhmedMiller2007} utilized geographic information systems and multidimensional scaling technique to implement the time-space maps of automobiles.
Kurkcu et al.~\cite{KurkcuMirandaOzbay2017} developed a web-based tool to store, process, and visualize the public bus data, thus making the analysis and decision-making process easier. 
Overbye et al.~\cite{OverbyeWeber2000} discussed different visualization techniques, i.e., line flow visualization, contouring bus and line data, and data aggregation.
Hock et al.~\cite{HockMcGuiness2018} presented a new visualization technique based on geographical information where estimated power system data were used to monitor and track anomalies in the power system.
McGee et al.~\cite{McGeeGhoniemMelancon2019} and Interdonato et al.~\cite{InterdonatoMagnaniPerna2020} provided overviews of existing multi-layer network visualization techniques, and summarized the existing  approaches to simplify the multi-layer network data.

The above research falls short in creating an interactive interface for transit and power distribution system operators to monitor both system constraints. 
To fill this gap, we present a visualization tool to showcase the bus transit and power distribution data and their interdependency interactively, informatively, and intuitively.

\section{Visualization Design and Observations}
\label{sec.vis}
A visualization tool is designed to illustrate the interdependent operations of bus transit and power distribution systems. 
The visualization interface is demonstrated in Fig.~\ref{fig.vis}. 
It is constructed for the operation of a test system based on the bus transit system of Park City, UT, together with a 33-bus power distribution system.
More information about the test system and its operation is provided in~\cite{bagherinezhad2020spatio}. 
The visualization tool is implemented in \textit{D3.js}. 
It can be viewed online at \url{https://usmart.ece.utah.edu/power-transit-vis/}, at the time of writing.

\subsection{Bus Transit and Power Distribution Systems}
\begin{figure*}[t]
\centering{
  \includegraphics[width=1\linewidth]{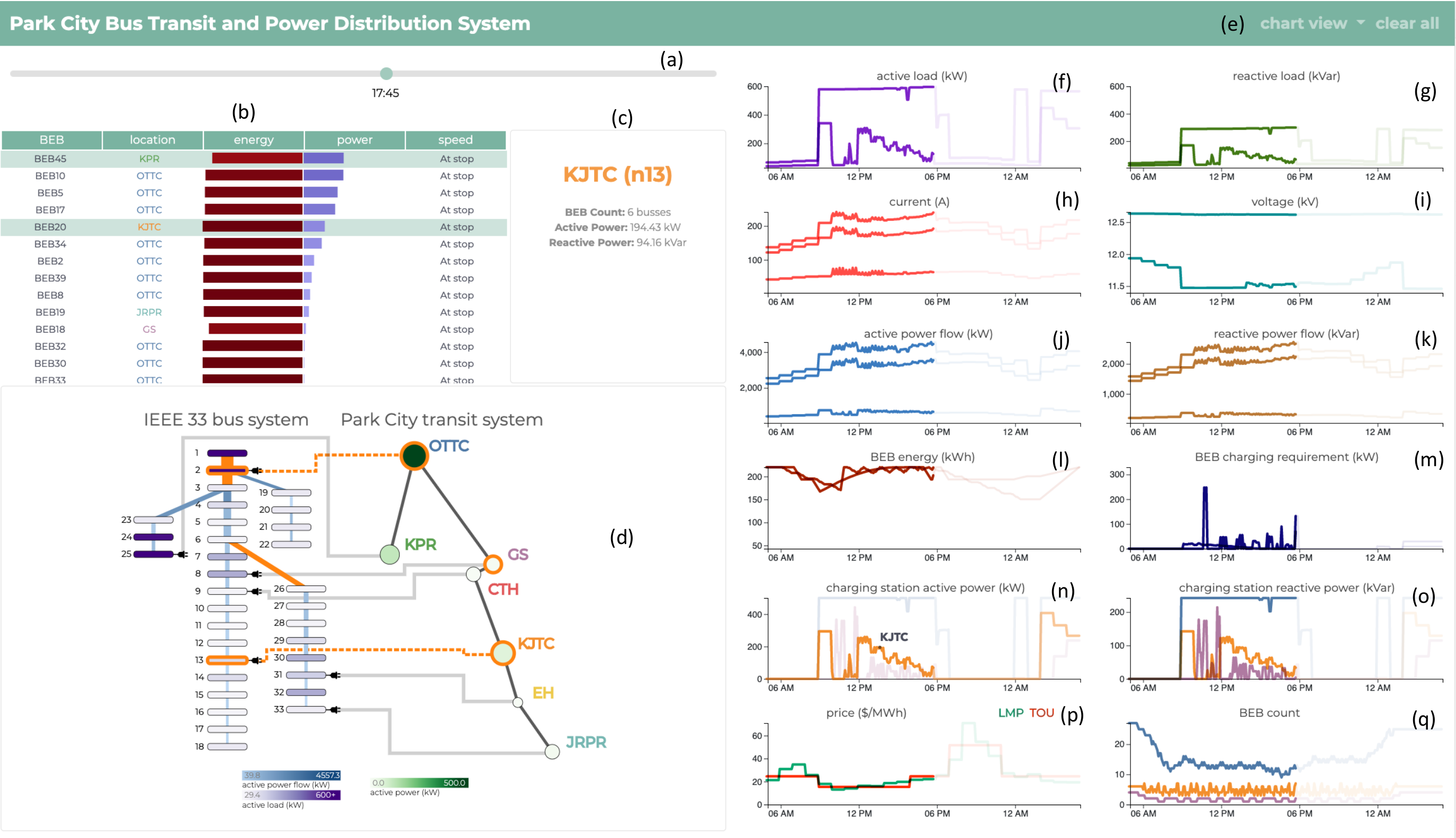}}
\vspace{-6mm}
  \caption{A visualization prototype that illustrates the interdependency of a battery  bus transit system and a power distribution system.}
  \label{fig.vis}
\vspace{-2mm} 
\end{figure*}

The interdependent systems are illustrated in Fig.~\ref{fig.vis}(d) left. 
The power distribution system comprises 33 nodes, where the first node is the substation that is connected to an upstream transmission network.
The power distribution nodes are denoted by purple rectangles; the intensity of the color represents the magnitude of the active load at a given time. 
The voltage, active load, and reactive load are data associated with the power distribution nodes. 
The power distribution nodes are connected to each other by blue lines, where the width of the lines represents the current, and the color intensity indicates the magnitude of their active power flow. 
The active power flow, reactive power flow, and current are data associated with the power distribution lines.

The test bus transit system comprises 9 routes with inter-station distances that range from 3 to 15.5 miles.
The battery bus transit system comprises 45 BEBs and 7 bus stops with charging infrastructure. 
In an effort to maintain simplicity, only the 7 bus stops with charging infrastructure (referred to as charging stations) are used in the visualization to denote the battery bus transit system (Fig.~\ref{fig.vis}(d) right).
The charging stations are arranged to mimic their actual geographical locations as seen from a bird's eye view. 
Each charging station node is represented by a circle, the radius of which indicates the number of BEBs that are currently at the station. 
Additionally, the circles are colored with varying shades of green to indicate the charging station active power. 
The charging station active power, reactive power, and the number of BEBs currently stopped at a station are part of the charging station data. 

The charging stations are connected to the power distribution nodes using gray lines.  
The data associated with these lines are a combination of the charging station data and their corresponding power distribution node data.
The entire interdependent system is located within a panel to enable panning and zooming, thus allowing the system operators to zoom in and focus on certain areas of the system.  

\subsection{Time and Data Panel}
The bus transit and power distribution data are available for 24 hours, spanning from 5am of the current day to 5am of the next day, in 5-minute time intervals. 
The system operator (a.k.a., the user) can change the current time via a slider shown in Fig.~\ref{fig.vis}(a), in which the visualization is  updated seamlessly as the time changes. 
In Fig.~\ref{fig.vis}(c), a data panel displays the current data values associated with a system component; it provides interactive data display (Fig.~\ref{fig.vis}(c)) via linked-views when an item in the table is hovered over in Fig.~\ref{fig.vis}(b), or when a user hovers over a line in the line charts in Figs.~\ref{fig.vis}(f-q).
For instance, hovering over a charging station in Figs.~\ref{fig.vis}(d)  would display the active and reactive power, and the number of parked BEBs in the data panel; and clicking on the charging station populates these data items to their corresponding line charts in Figs.~\ref{fig.vis}(f-q).
Additionally, the system operator can hover over lines in the charts across all time points, and the corresponding values appear in the data panel. 
The system operator can also adjust the time by clicking on the line charts.

\subsection{BEB Table}
The 45 BEBs in the bus transit system are visualized using a scrollable table, in which the data items can be sorted by clicking on the column headers in Fig.~\ref{fig.vis}(b).
This table lists BEBs' velocities and locations for a given time, and encodes the SOC of batteries and charging power requirements using diverging bars colored by the respective energy and power magnitude.  
The velocity of each BEB is dictated by the constant speed required to maintain the timetable of a bus transit system. 
The location of a BEB describes whether it is parked at a specific charging station or  \emph{on the road}. 
Additionally, by hovering over a BEB, the station that it most recently visited will  flash red, and the station it is currently headed toward will flash green.

\subsection{Price and Time of Use Tariff}
The locational marginal price (LMP) at which the power distribution system purchases electricity from the upstream transmission network, and the time-of-use (TOU) tariff at which the BEB charging requirements are priced, are respectively shown by green and red lines in Fig.~\ref{fig.vis}(p).  
The peak hours of TOU tariff for BEBs are during 6PM of the current day to 12 AM of the next day. 
According to the active power of a charging station, a transit system operator will opportunistically charge BEB batteries on road during off-peak and mid-peak hours (i.e.,~5 AM to 6PM of the current day, and 12AM to 5AM of the next day), see Fig.~\ref{fig.vis}(n).
This scheme results in reducing the energy cost of the battery transit system while maintaining power distribution and battery bus transit system constraints~\cite{bagherinezhad2020spatio}.

\subsection{Bus Transit and Power Distribution Systems Charts}
By default, the chart view shows charts of both power distribution and bus transit systems. 
However, to narrow one's focus, the chart view can be altered to show either charts relevant to the power distribution system or charts relevant to the bus transit system via the chart view dropdown button in Fig.~\ref{fig.vis}(e). 
The charts area (Figs.~\ref{fig.vis}(f-q)) displays line charts populated by operator selection of various components of the network and BEBs table.  
The user can select multiple components, in which each new selection will add new lines to the charts.
However, multiple lines on a chart make it difficult to distinguish lines from one another. 
This issue is circumvented by highlighting the line that is closest to the cursor as the system operator hovers over the line chart, while simultaneously diminishing the opacity of the other lines, see Fig.~\ref{fig.vis}(n). 
This feature allows the system operator to focus on an individual line while still seeing the trends of the other lines in the figure. 
To remove lines, the user can either click on a selected component, or select the \emph{clear all} button in the upper right corner.

\subsection{Visualizing Interdependency}
The interdependency of the battery bus transit components with the power distribution components is emphasized in several ways. 
First, it is emphasized through \emph{linked views}, in which hovering over a power distribution node that is linked with a charging station highlights the connecting line, and vice versa, as seen by the dashed orange line in Fig.~\ref{fig.vis}(d).
Second, hovering over a BEB highlights the corresponding station where it is currently located.
In addition to linked views and highlighting, an effective way to explore the interdependent nature of these systems is by clicking on one of the seven connecting lines. 
Upon clicking, both the power distribution node data and the charging station data are pushed to the line charts, facilitating the discovery of relationships between these data over time.
These highlighting methods ensure that the user is constantly aware of the inter-connectivity of bus transit and power distribution systems.

\subsection{Show, then Tell}
With 33 nodes and 32 lines in the power distribution system,  6 roads, 7 charging stations, and 45 BEBs in the bus transit system, as well as several unique, time-varying variables corresponding to each of these components, the primary challenge of our visualization tool is how to effectively   present such data in an informative and intuitive manner. 
Our main approach is to facilitate intelligent,  interactive selection of components for in-depth examination, which is accomplished in two parts. 
First, we present the system operator with high-level encodings designed to highlight potential components of interest with distinctive visual cues (i.e., color, position, and animation).
Second, we allow the system operator to examine these components further by clicking them and exploring the relevant line charts. 
This \emph{``show, then tell"} paradigm for visualization is particularly effective when dealing with large and complex data sets.

\section{Conclusion and Applications} 
\label{sec.conclusion}

We present a visualization tool to illustrate the interdependency of the battery bus transit and power distribution systems. 
The visualization tool effectively monitors transit and power distribution constraints (i.e., BEBs' SOC, electricity price, power flow, and voltage level) to determine the charging strategies of BEBs.
The tool provides visual cues for insights and discovery.
It also provides embedded chart viewing options to enable separate and focused visualizations of the two systems. 
Interactive visualization of the transit components pave the way for tailoring the spatio-temporal BEB charging demands in order to reduce the operation costs and maintain the quality of service. 
For instance, given the information on the SOC of BEBs, the available power capacity at charging stations, and the potential on-route charging stations along each BEB path, a transit system operator has the opportunity to alter the charging demands of BEBs spatially.
Moreover, the availability of the TOU tariff lets the bus transit operator opportunistically charge BEB batteries on the road during off-peak and mid-peak hours to reduce the energy cost in the bus transit operation. 

Furthermore, the visualization tool creates situational awareness of active and reactive power flow, spatio-temporal charging requirements of BEBs, current, and voltage level. 
These features allow power distribution system operators to change the course of action, for instance, to schedule distributed energy resources as needed to secure reliable operation.
In addition, the interdependent visualization tool creates the opportunity to co-optimize the operation of the bus transit and power distribution system and jointly solve and visualize the battery bus transit and power distribution constraints. 

\section*{Acknowledgement}
This work was partially supported by NSF IIS-1513616.

\bibliographystyle{IEEEtran}


\end{document}